\documentclass[preprintnumbers,article,amsmath,amssymb,floatfix,10pt,prd,twocolumn,
superscriptaddress,nofootinbib]{revtex4-2}
\usepackage{bm}
\usepackage{amsfonts}
\usepackage{latexsym}
\usepackage[latin1]{inputenc}
\usepackage{graphicx}
\usepackage{amsmath}
\usepackage{palatino}
\usepackage{mathpazo}
\usepackage{textcomp}
\usepackage{float}
\usepackage{booktabs}
\usepackage{bigints} 
\usepackage{dcolumn}
\usepackage{ragged2e}
\usepackage{hyperref}
\hypersetup{colorlinks,citecolor=blue}
\hypersetup{colorlinks=true,linkcolor=red,filecolor=magenta,    urlcolor=blue}
\usepackage{amsmath}
\usepackage{xcolor}
\usepackage{orcidlink}
\usepackage{epsfig}
\usepackage{subfigure}
\usepackage{commath}
\usepackage{mathtools}

\def\jnl@style{\it}
\def\aaref@jnl#1{{\jnl@style#1}}

\def\aaref@jnl#1{{\jnl@style#1}}

\def\aj{\aaref@jnl{AJ}}                   
\def\apj{\aaref@jnl{ApJ}}                 
\def\apjl{\aaref@jnl{ApJ}}                
\def\apjs{\aaref@jnl{ApJS}}               
\def\apss{\aaref@jnl{Ap\&SS}}             
\def\aap{\aaref@jnl{A\&A}}                
\def\aapr{\aaref@jnl{A\&A~Rev.}}          
\def\aaps{\aaref@jnl{A\&AS}}              
\def\mnras{\aaref@jnl{Mon.~Not.~Roy.~Astron.~Soc.}}             
\def\prd{\aaref@jnl{Phys.~Rev.~D}}        
\def\prc{\aaref@jnl{Phys.~Rev.~C}}  
\def\prl{\aaref@jnl{Phys.~Rev.~Lett.}}    
\def\qjras{\aaref@jnl{QJRAS}}             
\def\skytel{\aaref@jnl{S\&T}}             
\def\ssr{\aaref@jnl{Space~Sci.~Rev.}}     
\def\zap{\aaref@jnl{ZAp}}                 
\def\nat{\aaref@jnl{Nature}}              
\def\aplett{\aaref@jnl{Astrophys.~Lett.}} 
\def\apspr{\aaref@jnl{Astrophys.~Space~Phys.~Res.}} 
\def\physrep{\aaref@jnl{Phys.~Rep.}}      
\def\physscr{\aaref@jnl{Phys.~Scr}}       
\def\commat{\aaref@jnl{Comm.~Math.~Phys.}}              
\def\science{\aaref@jnl{Science}}               
\def\cqg{\aaref@jnl{Classical Quant.~Grav.}}            
\def\jpcs{\aaref@jnl{JPCS}}                                     
\def\ijmpd{\aaref@jnl{Int.~J.~Mod.~Phys.~D}}                    
\def\grg{\aaref@jnl{Gen.~Relat.~Gravit.}}               
\def\rpp{\aaref@jnl{Rep.~Prog.~Phys.}}          
\def\npa{\aaref@jnl{Nucl.~Phys.~A}}        
\def\lrr{\aaref@jnl{Living Rev.~Rel.}}                   
\def\jcap{\aaref@jnl{J.~Cosmology Astropart.~Phys.}}    
\def\rmp{\aaref@jnl{Rev.~Mod.~Phys.}}   
\def\epjc{\aaref@jnl{Eur.~Phys.~J.~C}} 
\def\plb{\aaref@jnl{~Phy.~Lett.~B}} 
\def\mpla{\aaref@jnl{Mod.~Phy.~Lett.~A}} 
\def\arxiv{\aaref@jnl{arxiv.org}}

\allowdisplaybreaks[1]

\begin{document}
\color{black}       
\title{Dehnen-type dark matter wormholes in the $f(\mathcal{R},\mathcal{L}_m,\mathcal{T})$ action}
\author{A. Errehymy\orcidlink{0000-0002-0253-3578}}
\email{abdelghani.errehymy@gmail.com}
\affiliation{Astrophysics Research Centre, School of Mathematics, Statistics and Computer Science, University of KwaZulu-Natal, Private Bag X54001, Durban 4000, South Africa}
\affiliation{Center for Theoretical Physics, Khazar University, 41 Mehseti Str., Baku, AZ1096, Azerbaijan}

\author{O. Donmez\orcidlink{0000-0001-9017-2452}}
\email{orhan.donmez@aum.edu.kw}
\affiliation{College of Engineering and Technology, American University of the Middle East, Egaila 54200, Kuwait}

\author{B. Turimov\orcidlink{0000-0003-1502-2053}}
\email{bturimov@astrin.uz}
\affiliation{Alfraganus University, Yukori Karakamish Str. 2a, 100190 Tashkent, Uzbekistan}
\affiliation{University of Tashkent for Applied Sciences, Gavhar Str. 1, Tashkent, 100149, Uzbekiston}
\affiliation{Shahrisabz State Pedagogical Institute, Shahrisabz Str. 10, Shahrisabz, 181301,Uzbekiston}

\author{K. Myrzakulov\orcidlink{0000-0002-4189-8596}}
\email{krmyrzakulov@gmail.com}
\affiliation{Department of General and Theoretical Physics, L.N. Gumilyov Eurasian National University, Astana 010008, Kazakhstan}

\author{N. Alessa\orcidlink{0000-0003-3283-4870}}
\email{naalessa@pnu.edu.sa}
\affiliation{Department of Mathematical Sciences, College of Science, Princess Nourah bint Abdulrahman University, P.O. Box 84428, Riyadh 11671, Saudi Arabia }

\author{A.-H. Abdel-Aty\orcidlink{0000-0002-6763-2569}}
\email{amabdelaty@ub.edu.sa}
\affiliation{Department of Physics, College of Science, University of Bisha, P.O. Box 344, Bisha 61922, Saudi Arabia}

\date{\today}
\begin{abstract}
We are exploring the possibility of traversable wormholes existing in a more realistic context. Specifically, we are looking at scenarios that don't rely on exotic factors, like having a mass shell at the throat or allowing particles and antiparticles to coexist without annihilation. To do this, we are constructing wormholes with double power-law density distributions, drawing inspiration from the Dehnen-type dark matter halo in the framework of generalized geometry-matter coupling gravity. Our investigation carefully considers the challenges of traversability and stability, as well as the roles of exotic matter, the exoticity parameter, and the anisotropy parameter. We have discovered solutions that describe asymmetric, asymptotically flat traversable wormholes, supported by a smooth metric and double power-law density distributions. These solutions successfully avoid the problems, giving us hope that such wormholes could actually exist in nature.
\end{abstract}

\maketitle


\textbf{Introduction:}
Wormholes are captivating concepts in physics that suggest connections between distant points in spacetime, and even between different universes \cite{Visser:1995yv}. While we haven't seen any wormholes yet, and there is still a lot of debate about whether they really exist or how they could form, the thought of one day creating a traversable wormhole or replicating one in a lab is incredibly exciting. This possibility has sparked a lot of interest among scientists and enthusiasts alike in recent years, as it taps into our natural curiosity about the universe and its hidden wonders.

For a wormhole to be safely traversable by humans, it needs to create gravitational repulsion, usually achieved by using exotic matter with negative kinetic energy that keeps the throat from collapsing. However, many examples of wormholes that do not rely on this kind of exotic matter often require significant changes to our current understanding of gravitational theory \cite{Bronnikov:2002rn, Bronnikov:2019sbx, Bronnikov:2003gx, Kanti:2011jz, Tomikawa:2014wxa, Bronnikov:2019sbx, Konoplya:2021hsm}. In fact, many of these theoretical wormholes are found to be unstable when subjected to small disturbances \cite{Gonzalez:2008wd, Cuyubamba:2018jdl}. Interestingly, tiny self-supported wormholes might be able to exist due to vacuum polarization effects happening around them \cite{Bronnikov:2018uje}. Additionally, the cylindrical wormhole solutions discovered in \cite{Bronnikov:2018uje} are noncompact and connect smoothly with an asymptotically flat spacetime. This raises an important question: Can asymptotically flat traversable wormholes exist as compact objects within Einstein's theory of gravity without needing phantom matter? If that is the case, it would mean that normal matter fields would have to violate the null energy conditions \cite{Morris:1988tu, Bolokhov:2021fil}. Before the discovery of wormhole solutions through the incorporation of Maxwell and two Dirac fields with their typical interactions, no solutions to Einstein's equations involving standard matter fields were known to support such wormholes, as shown by L\'{a}zquez-Salcedo and co-workers \cite{Blazquez-Salcedo:2020czn}, and include two main types. The first is an analytical solution describing a symmetric throat wormhole supported by massless, neutral fermions that are nonsymmetric and non-normalizable at infinity. The second solution, which is normalizable and obtained numerically, features symmetric configurations of the metric tensor and matter fields. This leads to some "exotic" properties, such as the need for a massive shell at the throat, the separation of fermion particles and antiparticles without annihilation, and discontinuities in differentiability, which ultimately make a consistent quantum description impossible.

The impact of surrounding matter becomes especially important in the context of gravitational lensing, particularly when dark matter is involved---a fundamental but still mysterious component of the cosmos. Although its presence has been inferred from observations such as galactic rotation curves and large-scale structure measurements \cite{Planck:2018vyg, Rubin:1970zza}, the true nature of dark matter remains unknown. Various theoretical models suggest that black holes could be embedded in dark matter-rich environments, potentially altering their gravitational lensing behavior \cite{Eiroa:2002mk, Xu:2018wow}. A promising method for representing dark matter halos involves the use of Dehnen-like profiles, which capture how matter density varies within galactic halos, especially near the central regions where the density can sharply increase \cite{Dehnen:1993uh, Navarro:1996gj}. These profiles are extensively employed in both simulations and observational studies of galaxy clusters, helping researchers explore how dark matter influences gravitational lensing phenomena.

Dark matter continues to be one of the most intriguing and mysterious elements in our understanding of the universe. Although it makes up roughly 27\% of the total cosmic energy density, it remains undetected through direct observation \cite{Bertone:2004pz, Planck:2015fie}. Its existence is primarily inferred through its gravitational influence---evident in the rotation curves of galaxies \cite{Rubin:1970zza}, the anisotropies in the cosmic microwave background \cite{Planck:2015fie}, and the distribution of matter on cosmic scales \cite{Springel:2005nw}. These observations highlight its fundamental role in shaping the formation and evolution of structures throughout the cosmos. Although we've made notable strides in understanding dark matter's gravitational role on cosmic scales, its true particle nature remains a mystery. Numerous candidates have been proposed over the years, such as axions, sterile neutrinos and weakly interacting massive particles \cite{Bertone:2018krk, Feng:2010gw}. 

Recently, growing interest has turned to the gravitational effects of dark matter halos---especially those surrounding galaxies and galaxy clusters---as a promising avenue to investigate its distribution and possible interactions. These extensive halos are commonly modeled using density profiles like the Navarro-Frenk-White profile \cite{Navarro:1996gj}, the Burkert profile \cite{Burkert:1995yz}, and the Dehnen profile \cite{Dehnen:1993uh}. Among various models, the Dehnen profile stands out as a flexible family of spherical mass distributions capable of capturing both the central and outer behavior of dark matter halos. Its generalized form accommodates both cuspy and cored density structures, making it well-suited for matching a wide range of observations---from dwarf galaxies to massive galaxy clusters \cite{Rubin:1970zza, Gohain:2024eer}. A specific case, known as the Dehnen$-(1,~4,~0)$ profile, supports a static, spherically symmetric spacetime \cite{Gohain:2024eer}, within which a Schwarzschild black hole can be embedded. The behavior of light (null geodesics) in such a system---where a Schwarzschild black hole is embedded within a Dehnen-type dark matter halo---has been thoroughly investigated in recent studies \cite{Gohain:2024eer,Jha:2024ltc}.

Here, we are interested in exploring whether traversable wormholes could exist in a more realistic setting. Specifically, we want to consider scenarios that don't involve the exotic factors previously mentioned, such as having a mass shell at the throat or allowing particles and antiparticles to coexist without annihilation. We are also investigating the construction of wormholes with double power-law density distributions, drawing insights from the Dehnen-type dark matter halo \cite{d1} in the $f(\mathcal{R},\mathcal{L}_m,\mathcal{T})$ action. Other notable recent works exploring astrophysical and cosmological phenomena can be found, particularly in the context of dark matter \cite{Maurya:2024jos, Errehymy:2024mlf, Errehymy:2024lhl, Errehymy:2023xpc, Errehymy:2023rnd}
 and $f(\mathcal{R},\mathcal{L}_m,\mathcal{T})$ gravity \cite{Tayde:2025yzq, Mohanty:2025mig, Loewer:2024lai, Errehymy:2025kzj, Errehymy:2025nzt}. In parallel, substantial progress has been made in exploring various astrophysical scenarios~\cite{Yousaf:2025tay, Yousaf:2024non, Khan:2024fuh, Khan:2024vsh, Yousaf:2024src}, offering valuable insights into the structure and behavior of self-gravitating fuzzy droplets, fuzzy black holes, and other compact objects with spherical symmetry~\cite{Aman:2025iye, Errehymy:2024tqr, Khokhar:2025ixd, Errehymy:2025haq, Yousaf:2025owb, Errehymy:2024qpu, Yousaf:2024kiv, Yousaf:2024rag}. These works have deepened our understanding of how matter interacts with gravity in some of the universe's most extreme environments. Alongside these developments, a growing body of research has focused on wormhole geometries~\cite{Lobo:2009ip, Bahamonde:2016ixz, Karakasis:2021tqx}, uncovering how factors such as anisotropic pressure, exotic matter, and modifications to gravity shape the conditions necessary for these fascinating structures to exist and remain stable.\\

\textbf{The $f(\mathcal{R},\mathcal{L}_m,\mathcal{T})$ gravity:}
In the study of compact objects, the $f(R, L_m, T)$ gravity theory---first introduced by Haghani and Harko~\cite{Haghani:2021fpx}---offers a compelling and versatile framework. This theory generalizes and unifies the models of $f(R, T)$ and $f(R, L_m)$ gravity, where $R$ is the Ricci scalar, $T$ is the trace of the energy-momentum tensor $T_{\mu\nu}$, and $L_m$ denotes the matter Lagrangian. In $f(R, L_m, T)$ gravity, the gravitational Lagrangian is given by an arbitrary function of these three quantities, written as  $L_{grav}=f(R, L_m,T)$. Assuming natural units where $G_N = c = 1$, the total action in this theory takes the form~\cite{Haghani:2021fpx}:
\begin{equation}\label{frl1}
    S=\frac{1}{16\pi}\int d^4xf(R,L_m, T)\sqrt{-g}+\int d^4xL_m \sqrt{-g}.
\end{equation}
In this framework, $g$ denotes the determinant of the spacetime metric $g_{\mu\nu}$. The energy-momentum tensor associated with the matter content is defined as
\begin{equation}
    T_{\mu \nu}=-\frac{2}{\sqrt{-g}}\frac{\delta(\sqrt{-g}L_m)}{\delta g^{\mu \nu}}.
\end{equation}

To determine how the trace of the energy-momentum tensor, $T = T^{\mu}_{\ \mu}$, varies with respect to the metric, we use the following relation:
\begin{eqnarray}
    \frac{\delta (g^{\alpha \beta}T_{\alpha \beta})}{\delta g^{\mu \nu}}=T_{\mu \nu}+\Theta_{\mu \nu},
\end{eqnarray}
where the tensor $\Theta_{\mu \nu}$ is given by
\begin{eqnarray}\label{theta}
    \Theta_{\mu \nu}\equiv g^{\alpha \beta}\frac{\delta T_{\alpha \beta}}{\delta g^{\mu \nu}}=L_mg_{\mu \nu}-2T_{\mu \nu}-2g^{\alpha \beta}\frac{\partial^{2}L_m}{\partial g^{\mu \nu} \partial g^{\alpha \beta}}.
\end{eqnarray} 

In this analysis, we adopt the common choice for the matter Lagrangian: $L_m = -\rho$, where $\rho$ denotes the energy density. We assume that $\rho$ depends only on the metric tensor $g_{\mu\nu}$ and not on its derivatives. In other words, $\rho$ is independent of terms like $\partial_\alpha g_{\mu\nu}$ or higher derivatives. This assumption allows us to ignore derivative-dependent contributions, specifically the second-order term in Eq.~\eqref{theta}, which therefore vanishes. As a result, $\Theta_{\mu \nu}$ simplifies to
\begin{equation}
    \Theta_{\mu \nu}=-2T_{\mu \nu}-\rho g_{\mu \nu}.
\end{equation}

The field equations of the $f(R, L_m, T)$ gravity theory are obtained by varying the action with respect to the inverse metric $g^{\mu\nu}$. This yields the following set of equations:
\begin{eqnarray}\label{frl2}
  &&8\pi T_{\mu\nu}+f_T\tau_{\mu\nu}\nonumber+\frac{1}{2}\left[f_{L_m}+2f_T\right]\left[T_{\mu\nu}-L_mg_{\mu\nu}\right]=-\frac{1}{2}fg_{\mu\nu} \nonumber\\
    &&+\left[R_{\mu\nu}+g_{\mu\nu}\Box-\nabla_\mu\nabla_\nu\right]f_R,
\end{eqnarray}
where $R_{\mu\nu}$ is the Ricci tensor and the derivative terms of the function $f$ are defined as:
\begin{eqnarray*}
f_R&\equiv& \frac{\partial f}{\partial R},~~
f_{L_m}\equiv \frac{\partial f}{\partial 
{L_m}},~~
f_{T}\equiv \frac{\partial f}{\partial T},~~ 
\tau_{\mu\nu}\equiv2g^{\alpha\beta}\frac{\partial^2L_m}{\partial g^{\mu\nu}\partial g^{\alpha\beta}}.
\end{eqnarray*}

The dynamics of gravity within the framework of $f(R, L_m, T)$ theories are highly sensitive to the specific form chosen for the function $f$. Each choice recovers or generalizes familiar gravitational models:
\begin{itemize}
    \item When $f(R, L_m, T) = f(R)$, the theory reduces to the well-known $f(R)$ gravity model.
    \item Choosing $f(R, L_m, T) = f(R, T)$ leads to the modified field equations of $f(R, T)$ gravity.
    \item If instead $f(R, L_m, T) = f(R, L_m)$, one obtains the formulation of $f(R, L_m)$ gravity.
    \item Finally, setting $f(R, L_m, T) = R$ recovers the standard Einstein field equations of GR:
\end{itemize}
\begin{eqnarray}
    8\pi T_{\mu\nu} &=& R_{\mu\nu}-\frac{1}{2}g_{\mu\nu}R. 
\end{eqnarray}

To explore a specific extension, consider a linear combination in the Lagrangian of the form:
\begin{equation}\label{LFRLm}
f(R, L_m, T) = R + \eta L_m + \chi T,
\end{equation}
where $\eta$ and $\chi$ are coupling constants, and the matter Lagrangian is taken to be $L_m = -\rho$. Under these assumptions, the generalized field equations simplify to:
\begin{equation}\label{frl3}
G_{\mu\nu}=\left[8\pi+\frac{\eta}{2}+\chi\right] T_{\mu \nu}+\frac{\chi}{2}\left[2\rho+T\right]g_{\mu\nu}.  
\end{equation}
This structure suggests a reformulation in terms of an effective stress-energy tensor:
 \begin{equation}
 G_{\mu \nu}=R_{\mu \nu}-\frac{1}{2}Rg_{\mu \nu}=\tilde{T}_{\mu \nu},
 \end{equation}
 where the effective tensor $\tilde{T}_{\mu \nu}$ combines conventional matter contributions with additional geometric sources:
 \begin{equation}
    \tilde{T}_{\mu \nu} = T^{(m)}_{\mu \nu} + \hat{T}^{(m)}_{\mu \nu},
\end{equation}
and the correction term is given by:
\begin{equation}
    \hat{T}^{(m)}_{\mu \nu} = \left(\frac{\eta}{2} + \chi\right) T_{\mu \nu} + \frac{\chi}{2}(-2 L_m + T) g_{\mu \nu}.
\end{equation}

To model the matter threading a wormhole geometry, we consider an anisotropic fluid characterized by the stress-energy tensor:
\begin{equation}\label{EMT}
    T_{\mu \nu}=(\rho+P_{t})u_{\mu}u_{\nu}+P_{t}g_{\mu \nu}+(P_{r}-P_{t})v_{\mu}v_{\nu}.
\end{equation}
where $u_\mu$ is a time-like unit vector and $v_\mu$ a space-like vector orthogonal to $u_\mu$. Under this framework, the energy-momentum tensor in diagonal form becomes:
\begin{equation}\label{cw8}
T^{\mu}_{\nu}=\text{diag}\left[-\rho, P_{r}, P_{t}, P_{t}\right],
\end{equation}
with $\rho$ representing the energy density, $P_r$ the radial pressure, and $P_t$ the transverse pressure. Given the form (\ref{LFRLm}), the trace of the energy-momentum tensor is computed as:
\begin{eqnarray}
T&=&-\rho +P_{r}+2P_{t},
\end{eqnarray}
and the standard matter tensor is retrieved via:
\begin{eqnarray}
8\pi T_{\mu \nu}&=&T^{(m)}_{\mu \nu}.
\end{eqnarray}

Here, the metric describes a static wormhole based on the Morris-Thorne solution \citep{Morris:1988cz}:
\begin{align}\label{cw1}
ds^2 = -e^{2\Phi(r)} dt^2 + \frac{dr^2}{1 - \frac{\hat{\psi}(r)}{r}} + r^2 (d\theta^2 + \sin^2 \theta \, d\phi^2),
\end{align}

where $\Phi$ represents the redshift function and $\hat{\psi}$ is the shape function.

To ensure that the wormhole behaves correctly at large distances, the redshift function must satisfy the condition:
\begin{align}
\lim_{r \rightarrow \infty} \Phi < \infty.
\end{align}

For the wormhole to be traversable at the throat located at $r_0$, it must meet specific criteria:
\begin{align}
\hat{\psi}(r_0) = r_0, \quad \hat{\psi}'(r_0) \leq 1.
\end{align}

Additionally, the shape function needs to follow these requirements away from the throat $r_0$: 
\begin{align}
\hat{\psi}' < \frac{\hat{\psi}}{r}, \quad \hat{\psi} < r.
\end{align}

The energy-momentum tensor associated with the wormhole is expressed as:
\begin{align}\label{cw8}
T^\mu_\nu = \text{diag}[-\rho, P_r, P_t, P_t],
\end{align}

where $\rho$ is the matter-energy density, $P_r$ is the radial pressure, and $P_t$ is the tangential pressure. This overview highlights the key components and conditions necessary for a viable wormhole structure.

Now, let's dive into the key non-zero components of the field equations that connect to the wormhole metric (\ref{cw1}) and the energy-momentum tensor (\ref{cw8}):
\begin{align}\label{14a}
    \frac{\hat{\psi}'}{r^2} &= \rho \left(\kappa - \frac{\chi}{2}\right) - \frac{\chi}{2} (P_r + 2 P_t)\,,\\\label{14b}
    -\frac{\hat{\psi}}{r^3} &= P_r \left(\kappa + \frac{\chi}{2}\right) + \frac{\chi}{2} (2 P_t + \rho)\,,\\\label{14c}
    \frac{1}{2 r^2}\left(\frac{\hat{\psi}}{r} - \hat{\psi}'\right) &= P_t (\kappa + \chi) + \frac{\chi}{2} (P_r + \rho)\,.
\end{align}
Here, we define $\kappa$ as $8 \pi + \frac{\eta}{2} + \chi$. We can express the quantities $\rho$, $P_r$, and $P_t$ in terms of $\hat{\psi}$. For instance, by combining Eqs. \eqref{14a} and \eqref{14b}, we can derive the sum $(\rho + P_r)$. This gives us the ability to find $P_t$ using Eq. \eqref{14c}. As a result, we can simplify our system of equations \eqref{14a}-\eqref{14c} to the following forms:
\begin{align}\label{15a}
   \hat{\psi}' &= \kappa r^2 \rho\,,\\\label{15b}
   -\hat{\psi} &= \kappa r^3 P_r\,,\\\label{15c}
   \hat{\psi} - \hat{\psi}' r &= 2 \kappa r^3 P_t \,.
\end{align}
When we look at our results alongside those from Moraes et al. \cite{Moraes:2024mxk}, we see that our final expressions align perfectly. This consistency really strengthens our conclusions.\\

\begin{figure*}
\centering
\includegraphics[width=4.35cm,height=4.35cm]{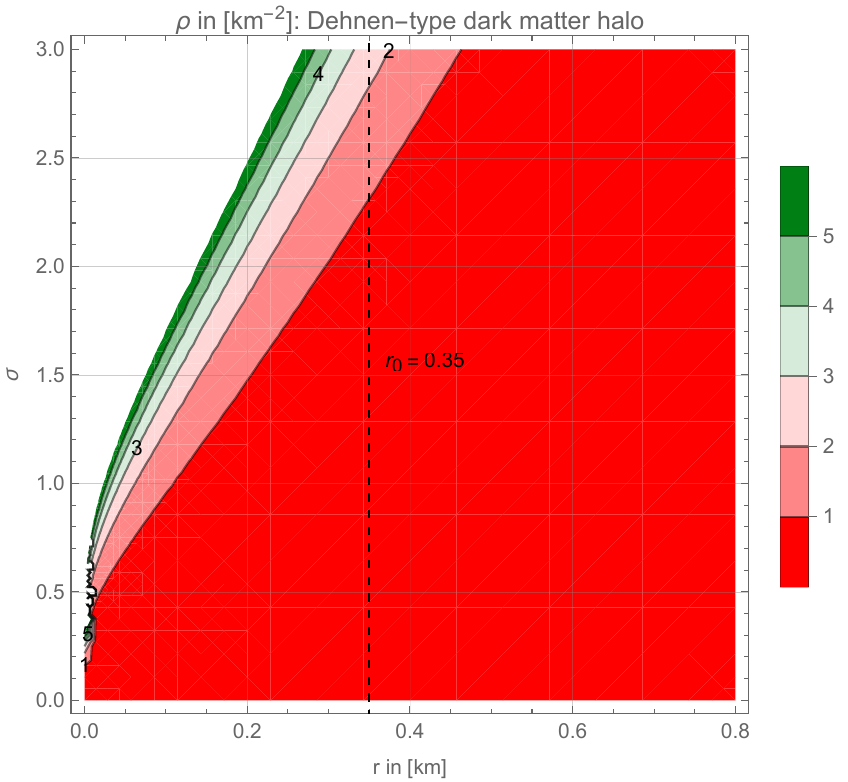}
\includegraphics[width=4.35cm,height=4.35cm]{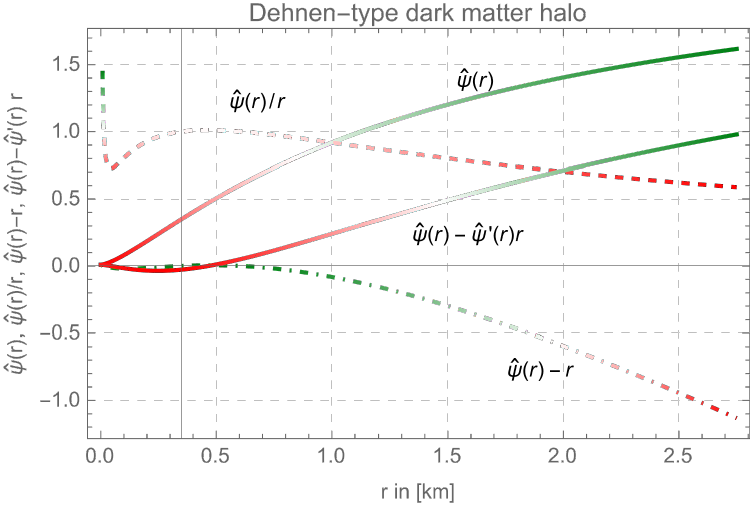}
\includegraphics[width=4.35cm,height=4.35cm]{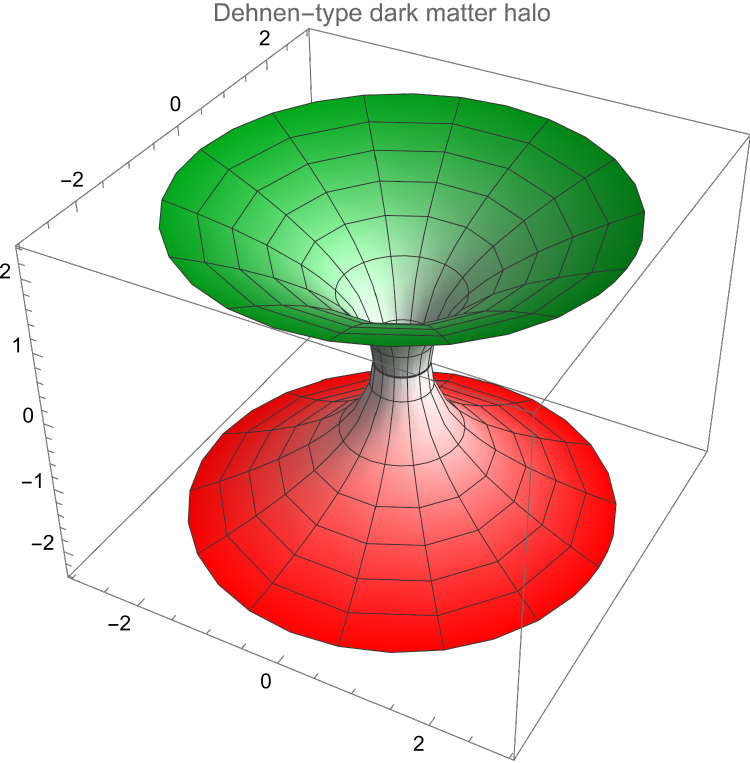}
\includegraphics[width=4.35cm,height=4.35cm]{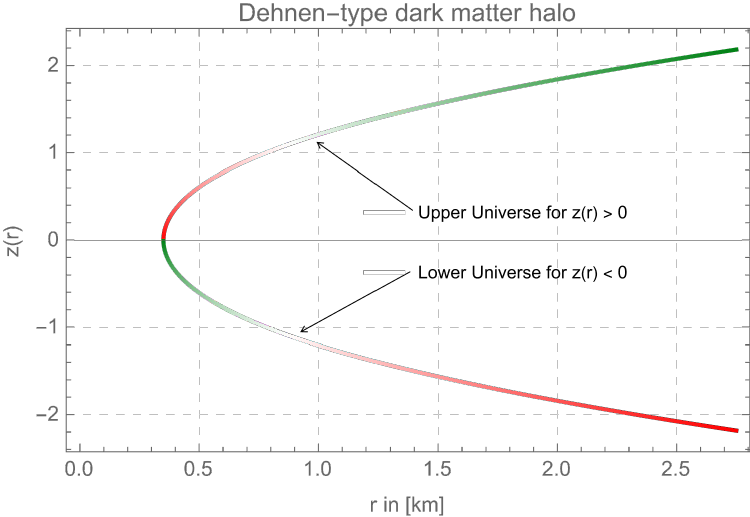}
\caption{The matter density and the conditions for the shape function of Dehnen-type dark matter wormholes are defined by the following key parameters: $\eta = 0.5, \quad \chi = 0.9, \quad \alpha = 1, \quad \beta = 4, \quad \rho_s = 0.15, \quad r_{s} = 0.99, \quad r_{0} = 0.35, \quad \sigma \in [0,~3]$. }\label{fig1}
\end{figure*}

\textbf{Specific solutions related to a double power-law profile: }
In this section, we will explore the density profile of the Dehnen dark matter halo. This halo is a specific example of a double power-law profile, defined by  \cite{d1}: 
\begin{equation}
\rho =\rho _{s}\left( \frac{r}{r_{s}}\right) ^{-\sigma }\left[ \left( 
\frac{r}{r_{s}}\right) ^{\alpha }+1\right] ^{\frac{\sigma -\beta }{\alpha }}.
\label{RhoCV}
\end{equation}%
Here, we will define $\rho_s$ and $r_s$ as the central density and radius of the halo. The parameter $\sigma$ specifies the variant of the profile, with values ranging from $[0, 3]$. For instance, when we use $\sigma = \frac{3}{2}$, it helps us fit the surface brightness profiles of elliptical galaxies, which closely resemble the de Vaucouleurs $r^{1/4}$ profile \cite{ds1}. 

Let us explore the intriguing connection between wormholes and the Dehnen dark matter halo by comparing \eqref{15a} with the density profile from \eqref{RhoCV}. This comparison is crucial for understanding how the density profile shapes a wormhole's form and stability, while examining the role of parameter $\sigma$ highlights the emptiness of these halos and its influence on wormhole properties, ultimately revealing fascinating links between the energy density and the unique features of wormholes.
\begin{equation}\label{23}
 \hat{\psi}'=\kappa \rho _{s}r^2\left( \frac{r}{r_{s}}\right) ^{-\sigma }\left[ \left( 
\frac{r}{r_{s}}\right) ^{\alpha }+1\right] ^{\frac{\sigma -\beta }{\alpha }}.
\end{equation}
Let us now integrate the shape function by applying the throat condition $\hat{\psi}(r_0) = r_0$, which simplifies our work by eliminating the integration constant and leads us to the following expression:
\begin{small}
\begin{align}\label{sf}
 \hat{\psi}&=-\frac{ \rho_s \kappa}{\sigma-3}\Biggl[r^3 \left(\frac{r}{r_s}\right)^{-\sigma} \, _2F_1\left(a,b;c;z_{r}\right)-r_0^3  \left(\frac{r_0}{r_s}\right)^{-\sigma} \,\nonumber\\& _2F_1\left(a,b;c;z_{r_0}\right)+r_0\Biggl],~~~~~~
\end{align}
\end{small}
with $_2F_1(a\equiv\frac{3-\sigma}{\alpha},b\equiv \frac{\beta-\sigma}{\alpha};c\equiv \frac{\alpha-\sigma+3}{\alpha};z\equiv z_{r}=-\left(\frac{r}{r_s}\right)^{\alpha},z_{r_0}=-\left(\frac{r_0}{r_s}\right)^{\alpha})$ being the hypergeometric function. The embedding surface $Z(r)$ is defined in an insightful way \cite{Morris:1988cz}
\begin{small}
\begin{align}
      Z(r)=\pm\bigintss_{\,r_0}^{\infty}\frac{1}{\sqrt{\frac{r}{\hat{\psi}}-1}}\text{d}r.
 \end{align}
 \end{small}
 The different signs show how the wormhole connects two distinct branches of spacetime.
 
 Let us derive the radial and transverse pressure components by simply substituting (\ref{sf}) into Eqs. (\ref{15b}) and (\ref{15c}):
\begin{small}
\begin{align}
    P_r(r)&= -\frac{r_0^3 \rho_s }{r^3(\sigma-3)} \left(\frac{r_0}{r_s}\right)^{-\sigma}\, _2F_1\left(a,b;c;z_{r_0}\right)-\frac{r_0}{\kappa r^3}\nonumber\\&+\frac{\rho_s  }{\sigma-3} \left(\frac{r}{r_s}\right)^{-\sigma}\, _2F_1\left(a,b;c;z_{r}\right),\label{p_r}
    \end{align}
    \begin{align}
    P_t(r)&=\frac{r_0^3 \rho_s \left(\frac{r_0}{r_s}\right)^{-\sigma} }{2r^3 (\sigma-3)}\, _2F_1\left(a,b;c;z_{r_0}\right)-\frac{\rho_s \left(\frac{r}{r_s}\right)^{-\sigma} }{2(\sigma-3)}\, _2F_1\left(a,b;c;z_{r}\right)\nonumber\\&-\frac{1}{2}\rho_s \left(\frac{r}{r_s}\right)^{-\sigma} \left(\left(\frac{r}{r_s}\right)^{\alpha}+1\right)^{\frac{\sigma-\beta}{\alpha}}+\frac{r_0}{2r^3 \kappa}.\label{p_t}
\end{align}
\end{small}\\

\begin{figure*}
\centering
\includegraphics[width=4.35cm,height=4.35cm]{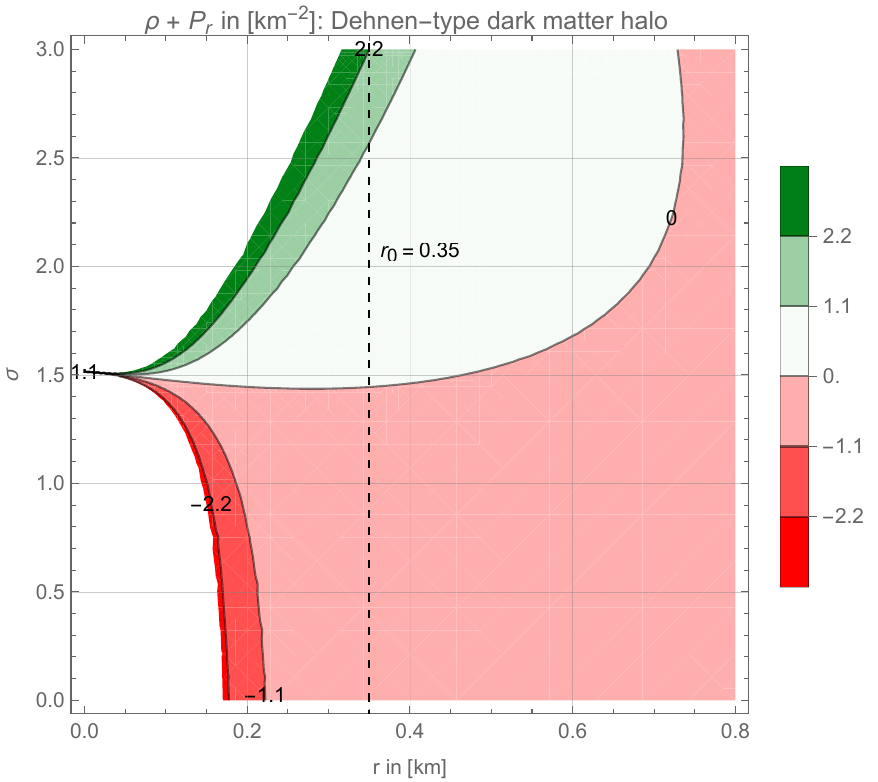}
\includegraphics[width=4.35cm,height=4.35cm]{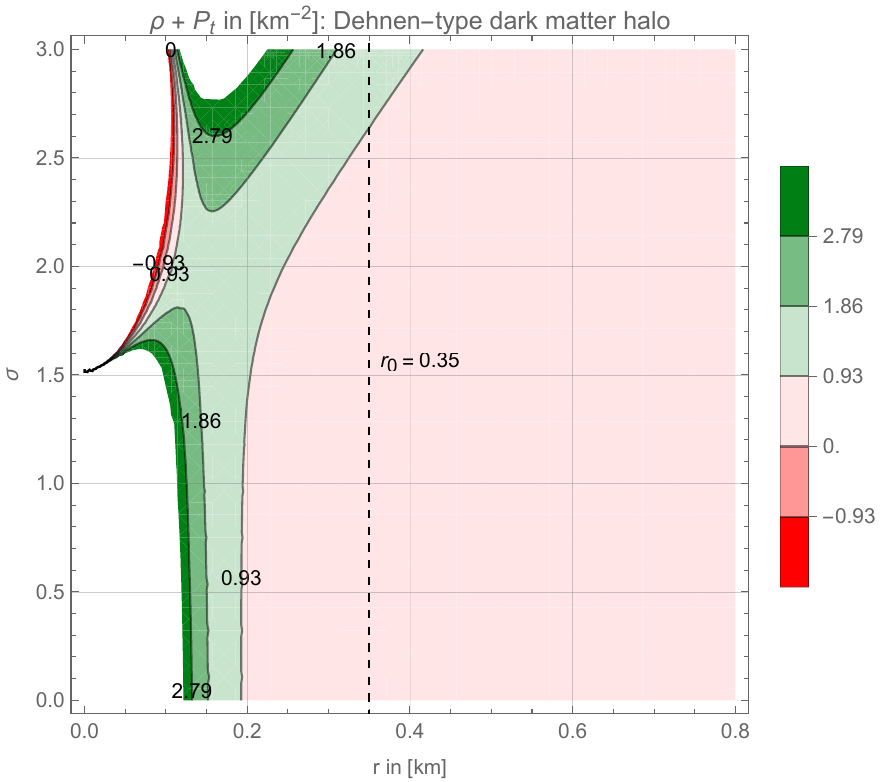}
\includegraphics[width=4.35cm,height=4.35cm]{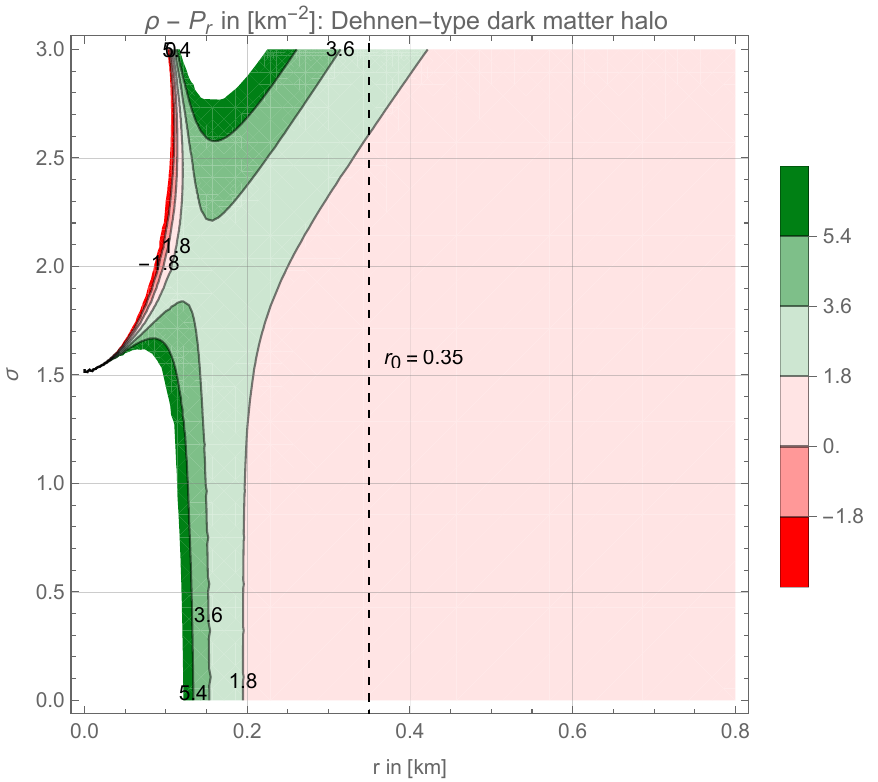}
\includegraphics[width=4.35cm,height=4.35cm]{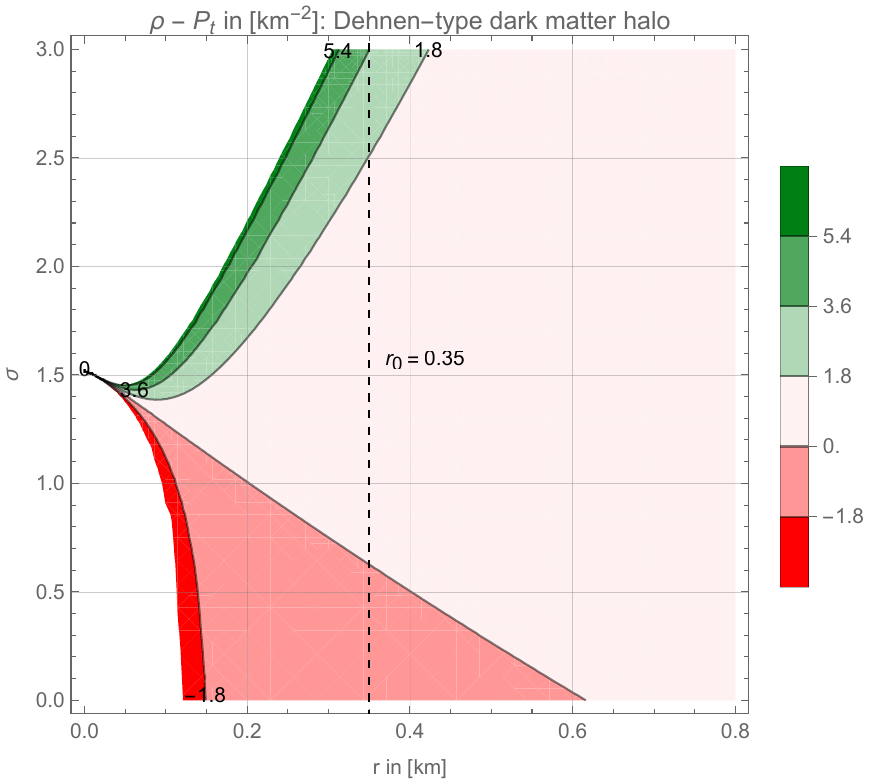}
\caption{The energy conditions for Dehnen-type dark matter wormholes are defined by the following key parameters: $\eta = 0.5, \quad \chi = 0.9, \quad \alpha = 1, \quad \beta = 4, \quad \rho_s = 0.15, \quad r_{s} = 0.99, \quad r_{0} = 0.35, \quad \sigma \in [0,~3]$. }\label{fig2}
\end{figure*}
\textbf{Tolman-Oppenheimer-Volkoff (TOV) Criteria:}
Grasping gravitational equilibrium in spherically symmetric spacetimes largely hinges on the TOV equation, which helps us evaluate stability in both GR and modified gravity scenarios \cite{Gorini}. This equation is flexible enough to account for anisotropic mass distributions, providing a richer framework for exploring these intriguing systems. The modified TOV equation can be expressed as:
\begin{align}\label{51}
\frac{\nu'}{2}(\rho+P_r)+\frac{dP_r}{dr}+\frac{2}{r}(P_r-P_t)=0.
\end{align}
Importantly, we noted previously that $\nu = 2\phi(r)$, which enhances our understanding of the various forces at play, including anisotropic, gravitational, and hydrostatic pressures:
\begin{align}\label{52}
F_h=-\frac{dP_r}{dr}, \quad F_g=-\frac{\nu'}{2}(\rho+P_r), \quad F_a=\frac{2}{r}(P_t-P_r).
\end{align}
We recognize that equilibrium is achieved when $F_h + F_g + F_a = 0$. Considering a constant redshift, $F_g$ becomes vanishing, leading to:
\begin{align}
\label{eq: eq}
    F_h+F_a=0.
\end{align}
In our study of linearly modified gravity, we find that the dynamics of the energy-momentum tensor present new terms \cite{Haghani:2021fpx}:
\begin{align}
    \nabla^\mu T_{\mu\nu}=-\frac{1}{\kappa}\Biggl[\Bigl(\frac{\eta}{2}+\chi\Bigr)\nabla_\nu\rho+\frac{1}{2}\Bigl(\chi\nabla_\nu T-\frac{\eta}{2}\nabla_\nu \rho\Bigr)\Biggr].
\end{align}
The force $F_m$ can be expressed as:
\begin{align}
    F_m = \frac{\chi}{2\kappa}(\rho' + P'_r + 2P'_t).
\end{align}
However, this force turns out to be zero everywhere because $\rho + P_r + 2P_t = 0$. Therefore, we will rely on \eqref{eq: eq} as the key condition for ensuring the stability of the solutions we are exploring.\\

\textbf{Exotic matter, exoticity parameter, and anisotropy parameter's influence on wormhole geometry:}
To understand the behavior of exotic materials near the neck of a wormhole, we consider the exoticity factor introduced by Lemos et al. in~\cite{Lemos:2003jb}, as a wormhole must be filled with exotic matter that violates the null energy condition and does not satisfy the weak energy condition or other related criteria
\begin{small}
\begin{align}
  \Omega_{Exoticity}=-\frac{\rho-P_r}{|\rho|}.
\end{align}
\end{small}
The presence of exotic matter near the throat of a wormhole is a fundamental requirement for maintaining the flare-out condition, which allows for traversability. This necessity is effectively captured by the exoticity factor, $\Omega_{Exoticity}$. A positive value of this factor indicates a dominance of exotic matter---i.e., matter that violates the null energy condition---near the throat. Conversely, a negative value implies that the exotic matter is diminishing and potentially transitioning into ordinary matter as one moves away from the throat. This behavior reflects a spatial variation in the nature of matter, as discussed by Lemos et al.~\cite{Lemos:2003jb} and Kim et al.~\cite{Kim:2003zb}.
The geometry and stability of wormholes are greatly affected by pressure anisotropy, which is crucial for figuring out if they can be traveled through. To explore the different attractive and repulsive shapes that wormholes can take, Cattoen et al.~\cite{Cattoen:2005he} and Lobo et al.~\cite{Lobo:2012qq} introduced a dimensionless anisotropy parameter as:
\begin{small}
\begin{align}
 \Delta=\frac{P_t-P_r}{\rho}.
\end{align}
\end{small}
When examining the throat of a wormhole, one encounters a fascinating pressure imbalance. On one hand, there is the radial pressure $P_r$, which acts inward or outward along the wormhole's central axis. On the other, the tangential pressure $P_t$, which acts sideways, perpendicular to that axis. The difference between these two pressures is called anisotropy, denoted by $\Delta = P_t - P_r$. This seemingly simple difference plays a key role in shaping the wormhole's structure and influencing whether it remains stable or collapses. If $\Delta > 0$, meaning the tangential pressure is greater than the radial one, the imbalance produces a kind of outward or repulsive effect. This can help counteract gravity's inward pull near the wormhole's throat, preventing collapse and maintaining the geometry needed for safe passage. In practical terms, this repulsion helps reduce destructive tidal forces and may even minimize the reliance on exotic matter---material with strange, negative-energy properties often considered necessary to keep wormholes open. In contrast, when $\Delta < 0$, the radial pressure wins out. This creates a net inward force that can destabilize the wormhole, intensify tidal effects, and increase the burden on exotic matter to keep the throat from pinching shut. Such a scenario raises serious concerns for traversability, as it could subject travelers to extreme physical forces and make the structure inherently less reliable.  Ultimately, the value and direction of $\Delta$ carry real physical meaning---they shape the wormhole's behavior, influence its stability, and determine if it can realistically be traveled through.

In $f(\mathcal{R}, \mathcal{L}_m, \mathcal{T})$ gravity, energy conditions are crucial for analyzing the causal and geodesic structure of space-time. A key approach to deriving these conditions involves the Raychaudhuri equations \cite{IV: RAY, Nojiri:2006ri, IV: CAP}, which characterize the dynamics of light-like, time-like, and space-like curves in the context of correspondence gravity. For scenarios involving an anisotropic fluid, the energy conditions relevant to $f(\mathcal{R}, \mathcal{L}_m, \mathcal{T})$ gravity are formulated as 
\begin{itemize}
    \item[(i)] NEC: $\rho + p_k \geq 0, \quad \forall k$,
    \item[(ii)] WEC: $\rho \geq 0, \quad \rho + p_k \geq 0, \quad \forall k$,
    \item[(iii)] DEC: $\rho \geq 0, \quad \rho \pm p_k \geq 0, \quad \forall k$,
    \item[(iv)] SEC: $ \rho + p_k \geq 0, \quad \rho + \sum_k p_k \geq 0, \quad \forall k$,
\end{itemize}
where  $k = r,~t$.

Let us delve deeper into the energy conditions by examining their expressions, taking advantage of the energy density $\rho$, the radial pressure $P_r$, and the tangential pressure $P_t$ in Eqs. \eqref{14a}-\eqref{14c}. By looking at these thermodynamic variables at a specific point, $r = r_0$, we can reformulate the NEC and explore what this means for the stability and possibility of wormholes in modified gravity scenarios.
\begin{figure*}[ht]
\centering
\includegraphics[width=4.35cm,height=4.35cm]{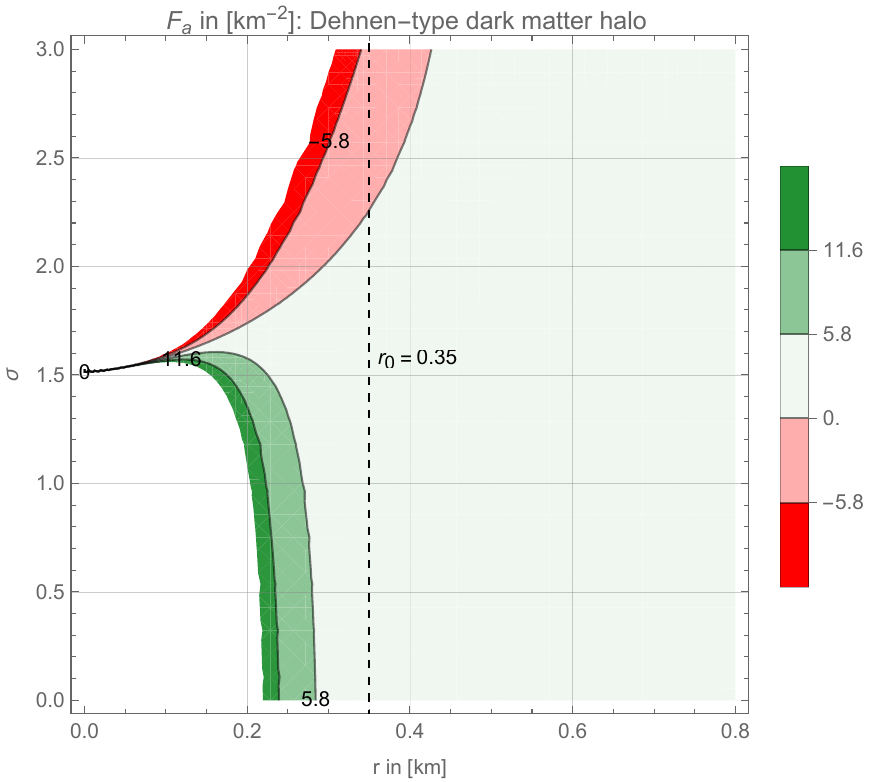}
\includegraphics[width=4.35cm,height=4.35cm]{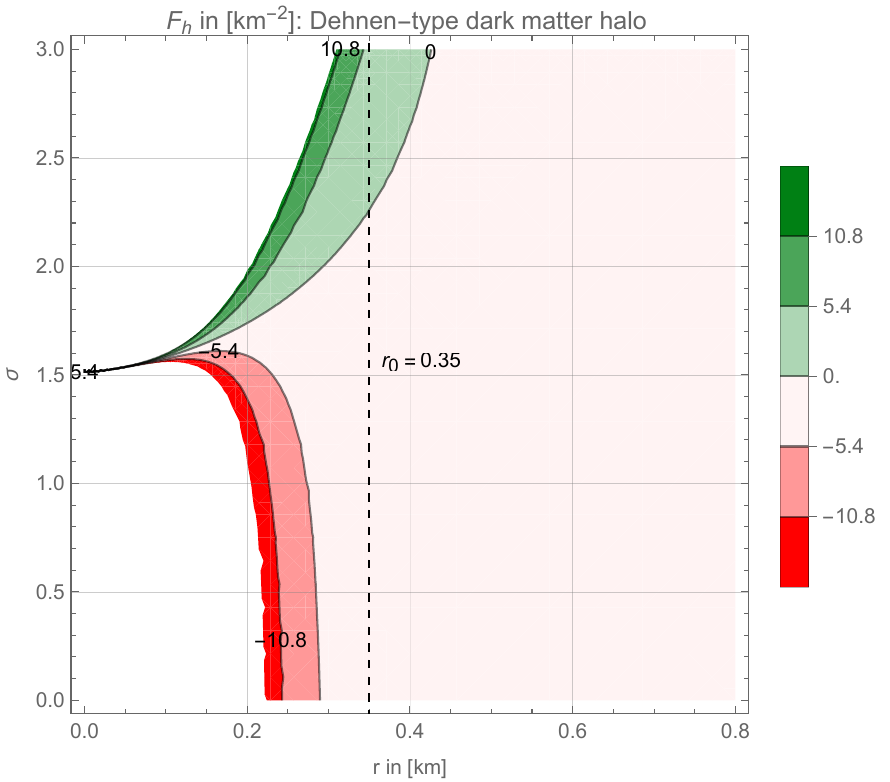}
\includegraphics[width=4.35cm,height=4.35cm]{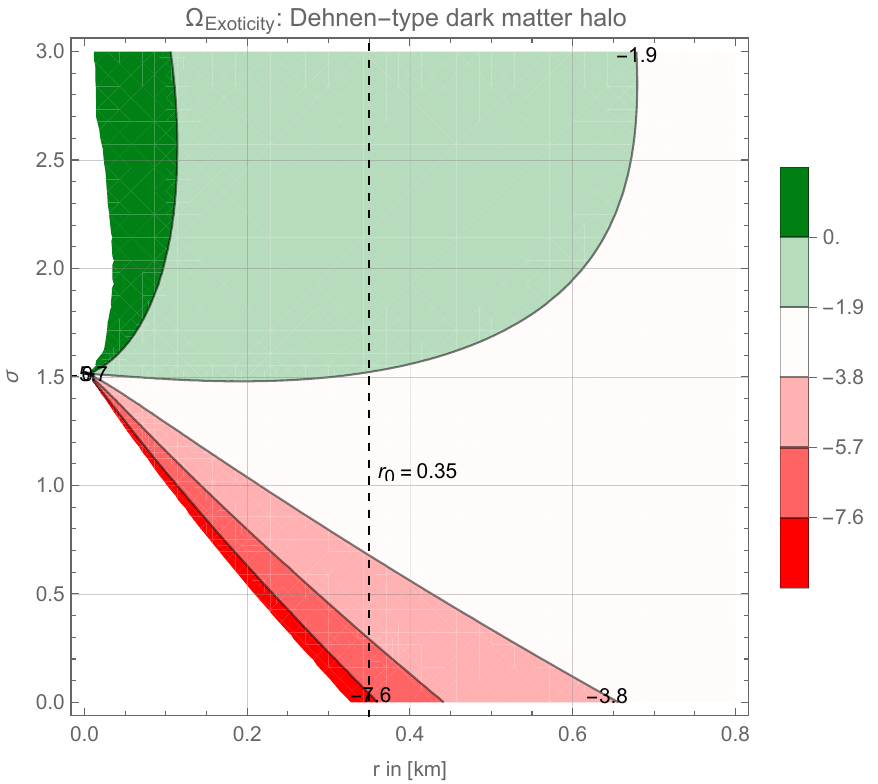}
\includegraphics[width=4.35cm,height=4.35cm]{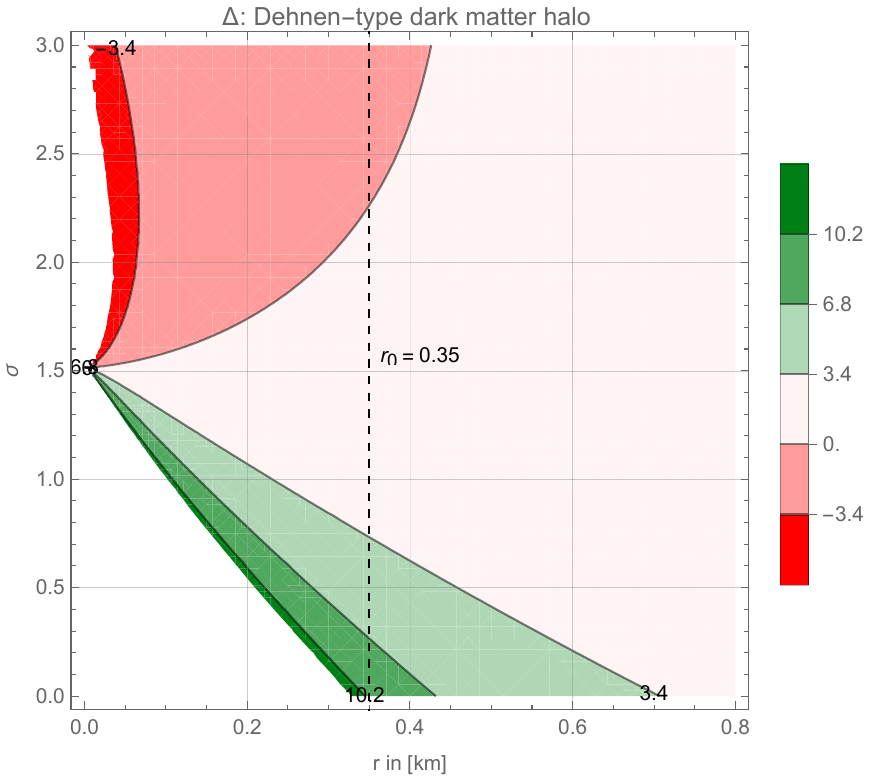}
\caption{ The anisotropic and hydrostatic forces, along with the exoticity and anisotropy parameters for Dehnen-type dark matter wormholes, are defined by the following key parameters: $\eta = 0.5, \quad \chi = 0.9, \quad \alpha = 1, \quad \beta = 4, \quad \rho_s = 0.15, \quad r_{s} = 0.99, \quad r_{0} = 0.35, \quad \sigma \in [0,~3]$. }\label{fig3}
\end{figure*}
\begin{align}
\begin{split}
\label{NECr}
 \rho+P_r\bigg\vert_{r=r_0}& = \rho_s \left(\frac{r_0}{r_s}\right)^{-\sigma} \left(\left(\frac{r_0}{r_s}\right)^{\alpha}+1\right)^{\frac{\sigma-\beta}{\alpha}}-\frac{1}{r_0^2 \kappa} \overset{!}{\geq}0\,\\
 \rho+P_t\bigg\vert_{r=r_0} &=   \frac{1}{2 r_0^2 \kappa}-\frac{\rho_s}{2} \left(\frac{r_0}{r_s}\right)^{-\sigma} \left(\left(\frac{r_0}{r_s}\right)^{\alpha}+1\right)^{\frac{\sigma-\beta}{\alpha}}\nonumber\\&+\rho_s \left(\frac{r_0}{r_s}\right)^{-\sigma} \left(\left(\frac{r_0}{r_s}\right)^{\alpha}+1\right)^{\frac{\sigma-\beta}{\alpha}}\overset{!}{\geq}0\,
        \end{split}
\end{align}
Now, let us turn our attention to the other energy conditions. A key point to keep in mind is that the WEC is satisfied whenever the NEC applies, especially since we consider $\rho$ to be positive. By carefully analyzing Eqs. \eqref{15a}-\eqref{15c}, we can derive the SEC, which is closely related to the NEC
\begin{align}
    \rho+P_r+2P_t=0.
\end{align}
We can explore whether a similar conclusion can be drawn for the DEC by expressing
\begin{align}
    \rho-|P_r|\bigg\vert_{r=r_0}&=\rho_s \left(\frac{r_0}{r_s}\right)^{-\sigma} \left(\left(\frac{r_0}{r_s}\right)^{\alpha}+1\right)^{\frac{\sigma-\beta}{\alpha}}--\frac{1}{r_0^2 \kappa}\\
     \rho-|P_t|\bigg\vert_{r=r_0}&= -\frac{1}{2r_0^2 \kappa}+-\frac{\rho_s}{2} \left(\frac{r_0}{r_s}\right)^{-\sigma} \left(\left(\frac{r_0}{r_s}\right)^{\alpha}+1\right)^{\frac{\sigma-\beta}{\alpha}}\nonumber\\&+\rho_s \left(\frac{r_0}{r_s}\right)^{-\sigma} \left(\left(\frac{r_0}{r_s}\right)^{\alpha}+1\right)^{\frac{\sigma-\beta}{\alpha}}.
\end{align}\\

\textbf{Concluding remarks: } \label{ch: VIII}
We have successfully explored the exciting possibility of traversable wormholes existing in a more realistic way. Specifically, we focused on scenarios that do not rely on exotic factors, like having a mass shell at the throat or allowing particles and antiparticles to coexist without annihilation. To do this, we constructed wormholes with double power-law density distributions, inspired by the Dehnen-type dark matter halo within the framework of generalized geometry-matter coupling gravity. For our analysis, we carefully selected specific parameter values: $\eta = 0.5, \quad \chi = 0.9, \quad \alpha = 1, \quad \beta = 4, \quad \rho_s = 0.15, \quad r_{s} = 0.99, \quad r_{0} = 0.35, \quad \sigma \in [0,~3]$. We are confident that our new shape function, inspired by the profile of the Dehnen-type dark matter halo, effectively reflects the core nature of wormholes. We examined key properties such as $\hat{\psi}$, $\frac{\hat{\psi}}{r}$, $\hat{\psi} - r$, and $\hat{\psi} - \hat{\psi}' r$ (see Fig. \ref{fig1}). Our findings show that when $r$ exceeds $r_0$, the value of $\hat{\psi} - r$ becomes negative, which means $\hat{\psi}/r < 1$. In simpler terms, as $r$ increases, the difference between $\hat{\psi}$ and $r$ shrinks. This behavior aligns perfectly with the flaring-out condition for $r \geq r_0$, ensuring that $\hat{\psi}'$ stays below 1. Moreover, both conditions $\hat{\psi}' < 1$ and $\hat{\psi}/r < 1$ hold for all $r$ greater than $r_0$. We also observe that $\hat{\psi}/r$ approaches 0 as $r$ goes to infinity, suggesting that spacetime appears flat at large distances. Additionally, our spacetime-embedding diagrams illustrate how mass is distributed according to the double power-law density distributions inspired by the Dehnen-type dark matter halo (see Fig. \ref{fig1} for more details). These visuals provide a captivating look into the geometric structure of the wormhole. When we analyze the energy density in this spacetime setup, we find that it remains positive across all examined parameters. This consistency indicates that the system effectively avoids any unphysical outcomes. It's interesting to note that variations in the parameter $\sigma$ can shift gravitational interactions, which in turn impact the energy conditions. Even when constrained between 0 and 3, the energy density stays positive, suggesting that the system is not only plausible, but also aligns well with standard energy conditions (see Fig. \ref{fig1}).

We now explore a different setup regarding the energy conditions (see Fig. \ref{fig2}). The expression $\rho + P_r$ is violated for $0 < \sigma < 1.44$ but is satisfied for $1.44 < \sigma < 3$ at the throat. In contrast, $\rho + P_t$ holds at the throat and beyond. The radial NEC is satisfied in the range $1.44 < \sigma < 4$, indicating a partially positive energy flow in the radial direction. Meanwhile, the complete positivity of the NEC suggests an entirely positive energy flow in the tangential direction, implying the absence of exotic matter. Varying $\sigma$ influences the violation of the radial NEC, potentially leading to phenomena such as traversable wormholes. Thus, $\sigma$ is crucial for adjusting the strength of interactions in this gravitational theory. Furthermore, $\rho - P_r$ is completely positive in the throat and continues to hold radially. However, $\rho - P_t$ is violated for $0 < \sigma < 0.64$ in the throat but is satisfied for $0.64 < \sigma < 3$. Finally, the expression $\rho + P_r + 2P_t = 0$ indicates that the SEC is satisfied. While these conditions may still depend on the presence of exotic matter, introducing a parameter based on the double power-law density distributions inspired by the Dehnen-type dark matter halo helps mitigate these violations.

When we analyze the forces involved, we discover that the stability of the wormhole relies on the TOV condition, expressed as $F_a + F_h = 0$. This means that the anisotropic force $F_a$ and the hydrostatic force $F_h$ are perfectly balanced-equal in strength but acting in opposite directions (see Fig. \ref{fig3}). This balance is consistent with our expectations from the double power-law density distributions inspired by the Dehnen-type dark matter halo. Interestingly, the exoticity parameter (see Fig. \ref{fig3}) shows a negative trend at the throat, suggesting that there is no exotic matter present in the context of $f(\mathcal{R}, \mathcal{L}_m, \mathcal{T})$ gravity. This absence of exotic matter highlights a fascinating aspect of the wormhole's structure. We also see that when the anisotropy is negative ($\Delta < 0$), it creates an attractive geometry for values of $\sigma$ between $2.27$ and $3$. On the other hand, for values between $0$ and $2.27$, positive anisotropy ($\Delta > 0$) leads to a repulsive geometry in the throat, which becomes more attractive as we move away (see Fig. \ref{fig3}). Moreover, the interplay between anisotropic pressure and the violation of the NEC is crucial for supporting traversable wormholes with exotic matter. This relationship helps prevent collapse and allows for the existence of exotic spacetime geometries, providing valuable insights into realistic wormhole models within the framework of $f(\mathcal{R}, \mathcal{L}_m, \mathcal{T})$ gravity.

It has been increasingly recognized that wormholes can closely resemble black holes in many key aspects. Damour and Solodukhin \cite{Damour:2007ap} pointed out that features such as quasi-normal modes, accretion dynamics, and even no-hair properties may be effectively mimicked by wormhole geometries. Cardoso et al. \cite{Cardoso:2016rao} demonstrated that wormholes with a thin shell of phantom matter can produce an initial ringdown phase virtually indistinguishable from that of black holes, with deviations only emerging at later times---especially relevant in light of LIGO's first detection of gravitational waves \cite{LIGOScientific:2016aoc}. Konoplya and Zhidenko \cite{Konoplya:2016hmd} further showed that the ringdown behavior of wormholes can either mirror or differ from black holes, depending on the specific model.

These developments underscore the importance of continued investigation into how wormholes might be observationally distinguished from black holes. For further insight, readers can consult works such as \cite{Bambi:2013nla, Ohgami:2015nra, Shaikh:2017zfl, Dai:2019mse} and for studies on accretion disk imaging around compact objects, see \cite{Luminet:1979nyg, Stuchlik:2010zz, Gyulchev:2019tvk, Tian:2019yhn}. Exploring wormholes from this observational viewpoint offers a promising path for future research.\\

\section*{Acknowledgments }
This research was funded by the Science Committee of the Ministry of Science and Higher Education of the Republic of Kazakhstan (Grant No. AP23487178). This work was supported by Princess Nourah bint Abdulrahman University Researchers Supporting Project number (PNURSP2025R59), Princess Nourah bint Abdulrahman University, Riyadh, Saudi Arabia. The authors are thankful to the Deanship of Graduate Studies and Scientific Research at University of Bisha for supporting this work through the Fast-Track Research Support Program. AE thanks the National Research Foundation of South Africa for the award of a postdoctoral fellowship.

\section*{Conflict Of Interest statement }
The authors declare that they have no known competing financial interests or personal relationships that could have appeared to influence the work reported in this paper.

\section*{Data Availability Statement} 
This manuscript has no associated data, or the data will not be deposited. (There is no observational data related to this article. The necessary calculations and graphic discussion can be made available
on request.)

\end{document}